\documentclass[10pt,a4paper]{article}
\usepackage[utf8]{inputenc}
\usepackage{amsmath}
\usepackage{amsfonts}
\usepackage{amssymb}
\usepackage{mathtools} 
\usepackage{float} 
\usepackage{tikz} 
\usepackage{pgfplots}
\usepackage{pgfplotstable}
\usepackage[numbers]{natbib}

\usetikzlibrary{patterns}
\usetikzlibrary{calc}
\usetikzlibrary{arrows,positioning,shapes,backgrounds,fit}  

\newcommand{\dt}[1][t]{\,\mathrm{d}{#1}}	
\newcommand{\bracket}[1]{\left( {#1} \right)}	
\newcommand{\der}[2][]{\frac{\mathrm{d} {#1} }{\mathrm{d}{#2}}}
\newcommand{\sder}[2][]{\frac{\mathrm{d}^2 {#1} }{\mathrm{d}{#2}^2}}
\newcommand{\partialder}[2][]{\frac{\partial {#1} }{\partial{#2}}}
\newcommand{\abs}[1]{\left| {#1} \right|} 		

\let\oldref\ref
\renewcommand{\ref}[1]{(\oldref{#1})}
\renewcommand{\epsilon}{\varepsilon}
\let\oldint\int
\renewcommand{\int}{\oldint\limits}

\begin{document}

\title{Fermat Principle and weak deflection angle from Lindstedt--Poincar\'e method}
\author{Joanna Piwnik\thanks{joanna.piwnik@edu.uni.lodz.pl}, Joanna Gonera \thanks{joanna.gonera@uni.lodz.pl}, Piotr Kosiński \thanks{piotr.kosinski@uni.lodz.pl}}
\date{}
\maketitle
\begin{center}
Faculty of Physics and Applied Informatics\\
University of Lodz, Poland
\end{center} 

\begin{abstract}
The Fermat principle is advocated to be a convenient tool to analyze the light propagation in a curved space time. It is shown that in the weak deflection regime the light ray trajectories can be systematically described by applying the Lindstedt--Poicar\'e method of solving perturbatively the nonlinear oscillation equations. The expansion in terms of inverse invariant impact parameter for Schwarzschild, Reissner--Nordström and Kerr (equatorial motion) metrics is described. The corresponding deflection angles are computed to the third order. Only algebraic operations are involved in the derivation; no integrations or Fourier expansion of elliptic functions are necessary. It is argued, that contrary to the naive perturbative expansion, the Lindstedt--Poincar\'e approach correctly represents the main properties of light propagation in asymptotic regime. At each step it preserves the periodicity of the relevant nonlinear oscillations of the inverse radial coordinate which allows to group the trajectories with the same invariant impact parameter into disjoint sets of ones generated by particular oscillations. Moreover, it allows for partial summation of perturbative expansion leading to the uniformly bounded approximations.    
\end{abstract}

\pgfplotstableread[col sep=comma]{trajectories_data.csv}\mydata
\pgfplotstableread[col sep=comma]{kat_od_mb.csv}\mydataa

\section{Introduction}
There are two long-ranged fields, which mediate interactions on the macroscopic, classical level: the electromagnetic field and the gravitational one; in fact, it seems that under some quite mild assumptions concerning quantum description of long-ranged relativistically covariant interactions these are the only fields that can appear on the classical level \cite{1}. It is, therefore, crucial to study their mutual interaction as well as the selfinteraction of gravitational field. The former concerns the propagation of electromagnetic field in gravitational background, the latter -- the propagation of gravitational waves. 

The study of the influence of gravity on light propagation is important for two reasons. First, with the incredible development of observation techniques we are faced with multitude of new phenomena like, for example, Einstein rings \cite{2}. Their correct interpretation demands good understanding of the theoretical aspects of light propagation in gravitational field. 

Second, there are many open questions concerning the gravity and the structure of Universe: the nature of dark matter and dark energy, the quantization of gravity etc. It may appear that the classical Einstein theory of gravitational field should be modified. Various such modifications, more or less ad hoc, have been proposed. In order to select the most promising ones the constraints following from the experiments and observations must be taken into account. The gravitational deflection and lensing remains one of the most indispensable tools to analyse such constraints.

Much effort has been devoted in last decades to the study of various aspects of light propagation in the presence of gravitational field. The general properties of the geometry of light rays in the presence of black holes were discussed in \cite{3, 4, 5, 6, 7, 8, 9, 10, 11, 12, 13, 14, 15, 16, 17, 18}. 

Historically, the simplest case of Schwarzschild metric has been considered by Darwin \cite{19}, \cite{20}, who gave the exact formula for deflection angle. The light propagation in the background of Schwarzschild black hole has been further discussed in \cite{21, 22, 23, 24, 25, 26, 27, 28, 29}.

The results obtained in this case can be extended in rather straightforward way to other spherically symmetric configurations like Reissner--Nordström metric \cite{30, 31, 32, 16}. The case of axially symmetric metrics is more complicated; typically it simplifies for equatorial motion \cite{34, 35, 36, 37, 38, 39, 40}. 
Asymptotically non-flat metric and naked singularities were also considered in \cite{40}, \cite{41} and \cite{42}, respectively.

The analysis of light propagation in gravitational field relies on solving the geodesic equation for null geodesics. Once the affine parameter is given the geodesic equation is derived from the Lagrangian quadratic in velocities (derivatives with respect to affine parameter). We are dealing with Lagrangian system with four degrees of freedom. The resulting dynamics may be integrable or not depending on the symmetry degree of the background metric; for example, in the static, spherically symmetric metric the geodesic equations describe the dynamical system which is integrable in the Arnold--Liouville sense. In general, the lower is the symmetry of the metric, the more difficult is to find the analytical explicit solution. Sometimes only some specific solutions are available; in the axially symmetric case, for example, only equatorial motions are easily analytically manageable (although nonequatorial motion is still integrable \cite{53}). 

Among various physically relevant configurations of gravitational field the constant (both static and stationary) fields play an important role. The geodesic equations may be then replaced by the Fermat principle \cite{43}. It has two advantages; first, we don’t have to consider any longer the affine parameter which has no direct physical meaning; actually, it is defined up to an affine transformation. Second, when solving the geodesic equations we have to select the null geodesics by imposing the additional constraint on the initial values; on the contrary, Fermat principle deals directly with null geodesics. Moreover, Fermat principle can be put in the Lagrangian form, with only two degrees of freedom involved.

In the present paper we argue that the Fermat principle provides a simple and straightforward framework for describing several properties of the light propagation in various, physically interesting space-times. In particular, we are interested in the weak  deflection limit. Then, at least in the cases of metrics exhibiting high degree of symmetry, the relevant equations become that of non-linear oscillations. The latter can be solved perturbatively by applying the Lindstedt--Poincar\'e method \cite{43, 44, 42}. It consists in rearranging the perturbation expansion in such a way as to avoid the occurrence of resonant terms, providing thereby a uniform successive approximation to the exact trajectories. We show that this approach allows for obtaining a systematic expansions of deflection angle as a power series in the inverse of invariant impact parameter by performing only algebraic operations; neither any integration nor Fourier expansion of elliptic functions are necessary. Moreover, some contributions to the relevant expansions can be interpreted in terms of frequency--amplitude relations typical for non-linear oscillations. For the spherically symmetric case the Lindstedt--Poincar\'e algorithm is directly applicable. In the less symmetric case of Kerr metric it must be slightly modified while keeping the main idea of eliminating the secular terms intact.

The paper is organized as follows. In section 2 we remind the Fermat principle for constant (both static and stationary) gravitational field. Sections 3 and 4 are devoted to the spherically and axially symmetric gravitational fields, respectively. We re-derive the exact expressions for light trajectories (in the equatorial plane in the axially symmetric case) using Fermat principle and elementary Lagrangian mechanics with time replaced by the azimuthal angle. Section 5 is devoted to the study of asymptotic trajectories with the help of Lindstedt--Poincar\'e method, while Section 6 describes the advantages of the latter.

Finally, some conclusions are presented in section 6. The Lindstedt--Poincar\'e method is briefly sketched in Appendix.

\section{The Fermat principle}
Let
\begin{equation}
\dt[s]^2 = g_{\mu\nu}(x) \dt[x]^\mu \dt[x]^\nu
\end{equation}
be the metric of the space-time under consideration. We will be dealing with a constant gravitational field, i.e. the one admitting the reference frame in which all components $g_{\mu \nu} (x)$ of the metric tensor do not depend on the time coordinate $x^0$. Furthermore, the metric is called static (stationary) if $g_{0k}=0$ for $k = 1,2,3$ ($g_{0k} \neq 0$ for some $k$).

The light propagation in a constant gravitational field is described by the Fermat principle. The latter may be formulated as follows (cf. \cite{43} where the Fermat principle is derived from the equivalence principle): from the light cone condition
\begin{equation}\label{2}
\dt[s]^2 = 0
\end{equation}
one computes $\dt[x]^0$ and defines the action functional
\begin{equation} \label{3}
I = \int \dt[x]^0.
\end{equation}
Then the Fermat principle takes the form
\begin{equation}\label{4}
\delta I = 0,
\end{equation}
with fixed endpoints, and involves three variables $x^k$, $k=1,2,3$. One can, at least locally, choose one of them as evolution parameters and solve the resulting Lagrange equations. Alternatively, all variables $x^k$ can be considered on equal footing yielding the constrained system.

Explicitly, the action integral $I$ has the form \cite{43}
\begin{equation}
I = \int \bracket{\frac{\dt[l]}{\sqrt{g_{00}}} - \frac{g_{0k} \dt[x]^k}{g_{00}}}
\end{equation}
where $\dt[l]$ is the length element of spatial distance.

Once the spatial trajectory of light is determined from \ref{4} one can find from \ref{3} the time coordinate. It is, basically, equal to the action $I$ computed along the trajectory. Once the Fermat principle is put in the Lagrangian form all tools of analytical mechanics become available.

\section{Spherically symmetric metric}
Let us start with the simplest case of the spherically symmetric metric described by the line element 
\begin{equation} \label{6}
\dt[s]^2 = B(r) \dt[t]^2 - A(r) \dt[r]^2 - r^2 D(r) \bracket{\dt[\theta]^2 + \sin^2\theta \dt[\phi]^2}.
\end{equation}
Assuming asymptotic flatness one finds $A(r)$, $B(r)$, $D(r) \xrightarrow[r \to \infty]{} 1$. Due to the spherical symmetry we can assume from the very beginning that the motion takes place in $\theta = \frac{\pi}{2}$ plane. Then \ref{6} reduces to 
\begin{equation}\label{7}
\dt[s]^2 = B(r) \dt[t]^2 - A(r) \dt[r]^2 - r^2 D(r) \dt[\phi]^2.
\end{equation}
The important quantity characterizing the trajectory is the invariant impact parameter. To define it let us remind that, given an affine parameter $\sigma$, one derives the geodesic equations from the variational principle
\begin{equation}
\delta \int L \dt[\sigma] = 0
\end{equation} 
where
\begin{align*}
L &\equiv \frac{1}{2} g_{\mu\nu}(x) \der[x^\mu]{\sigma} \der[x^\nu]{\sigma}\\
&= \frac{1}{2} \bracket{B(r) \bracket{\der[t]{\sigma}}^2 - A(r) \bracket{\der[r]{\sigma}}^2 - r^2 D(r) \bracket{\der[\phi]{\sigma}}^2}
\end{align*}
$L$ does not depend explicitly on $t$ and $\phi$, implying the conservation of the corresponding momenta
\begin{align}
E &\equiv p_t = \partialder[L]{\dot{t}} = B(r) \dot{t}, &\dot{t} \equiv \der[t]{\sigma};\\
J &\equiv p_\phi = \partialder[L]{\dot{\phi}} = -r^2 D(r) \dot{\phi}, &\dot{\phi} \equiv \der[\phi]{\sigma}.
\end{align}
The affine parameter $\sigma$ is defined up to an affine transformation, $\sigma \to a \sigma + b$; the latter implies rescaling $E \to E/a$, $J \to J/a$. 
Therefore, the physically relevant parameter, the invariant impact parameter, may be defined as
\begin{equation} \label{11}
b = \frac{\abs{J}}{E} = \frac{r^2 D(r)}{B(r)} \abs{\der[\phi]{t}}.
\end{equation}
Note that the dynamics defined by $L$ and the evolution parameter $\sigma$ is integrable in the Arnold--Liouville sense, possessing three Poisson-commuting integrals of motion $ \mathcal{H}_0 = L$, $p_t$ and $p_\phi$.

Now, we can write out the variational principle describing the Fermat law. Computing $\dt$ from \ref{2} and \ref{7} we find that the Fermat principle takes the form
\begin{equation}
\delta \int \sqrt{\frac{A(r)}{B(r)}\dt[r]^2 + \frac{r^2 D(r)}{B(r)} \dt[\phi]^2} = 
\delta \int \sqrt{\frac{A(r)}{B(r)}\bracket{\der[r]{\phi}}^2 + \frac{r^2 D(r)}{B(r)}} \dt[\phi] = 0.
\end{equation}
Therefore, the trajectory $r = r(\phi)$ can be read off from the Lagrange equation resulting from the Lagrangian
\begin{equation}\label{13}
\mathfrak{L} = \sqrt{\frac{A(r)}{B(r)}\bracket{\der[r]{\phi}}^2 + \frac{r^2 D(r)}{B(r)}}
\end{equation}
with $\phi$ replacing the time (evolution) parameter. $\mathfrak{L}$ does not depend explicitly on $\phi$. As a result the "energy"
\begin{equation}
\mathcal{E} \equiv \dot{r} \partialder[\mathfrak{L}]{\dot{r}} - \mathfrak{L},\ \dot{r} \equiv \der[r]{\phi}
\end{equation}
is a constant of motion. Explicitly,
\begin{equation}\label{15}
\mathcal{E} = \frac{-r^2 \frac{D(r)}{B(r)}}{\sqrt{\frac{A(r)}{B(r)}\dot{r}^2 + \frac{r^2 D(r)}{B(r)}}}
\end{equation}
Now, $\dt[t] = \mathfrak{L} \dt[\phi]$ yields
\begin{equation}\label{16}
b = \abs{\mathcal{E}}
\end{equation}
Equations \ref{15} and \ref{16} imply
\begin{equation}\label{17}
\frac{A(r) B(r)}{D^2(r)} \frac{\dot{r}^2}{r^4} + \frac{B(r)}{r^2 D(r)} = \frac{1}{b^2}.
\end{equation}
By taking derivative with respect to $\phi$ of both sides of eq.\ref{17} one obtains the relevant Lagrange equation.

Eq.\ref{17} allows to derive an explicit expression for the deflection angle. If $r_0$ denotes the distance of the closest approach of the light to the center, $\left. \dot{r} \right|_{r_0} = 0$, then
\begin{equation} \label{18}
\frac{1}{b^2} = \frac{B(r_0)}{r_0^2 D(r_0)}.
\end{equation}
Eqs.\ref{17} and \ref{18} yield immediately the deflection angle (cf. Fig.\oldref{fig3} below) (see \cite{44} for the $D(r) \equiv 1$ case and \cite{47a} for the general one)

\begin{figure}[H] 
\centering
\begin{tikzpicture}[scale=1]
\begin{scope}[xscale = 1,yscale = 1]
\draw[->] (0, 0) -- (6, 0) node[below] {$\phi=0$};
\draw[dashed] (0, 0) -- (-2, 3);
\draw[dashed] (0, 0) -- (-2, -3);
\draw[dashed] (2.66, -1) -- (0, 3);
\draw[dashed] (2.66, 1) -- (0, -3);
\draw[dashed] (2.3,0.5) arc [start angle=60, end angle=116, radius=0.7cm];
\node at (2,0.1)[above] {$\delta$};
\draw[dashed] (0.5,0) arc [start angle=0, end angle=116, radius=0.7cm];
\node at (0.1,0.1)[above] {$\phi_\infty$};
\draw (0,3) to[out=-56, in=56, looseness=1.6] (0,-3);
\end{scope}
\end{tikzpicture}
\caption{The geometry of deflection angle.} \label{fig3}
\end{figure}

\begin{equation}\label{19}
\delta = 2 \int_{r_0}^{\infty} \sqrt{\frac{A(r)}{D(r)} \bracket{\frac{D(r)}{B(r)} \frac{B(r_0)}{D(r_0)}\bracket{\frac{r}{r_0}}^2 -1}^{-1}} \frac{\dt[r]}{r} - \pi
\end{equation}
which allows us to compute $\delta$ in terms of $b$.

As a particular case of spherically symmetric case consider the metric describing space-time  corresponding to electrically neutral, compact, spherically symmetric mass distribution, i.e. the Schwarzschild metric. It is given by eq.\ref{7} with 
\begin{equation}
B(r) = A^{-1} (r) = 1 - \frac{2m}{r}, D(r) = 1,
\end{equation}
$m$ being the mass parameter. Then the Lagrangian \ref{13} takes the form
\begin{equation}
\mathfrak{L} = \sqrt{\frac{1}{B^2(r)} \bracket{\der[r]{\phi}}^2 + \frac{r^2}{B(r)}}
\end{equation}
while
\begin{equation} \label{22}
\mathcal{E} = \frac{- r^2}{\sqrt{\dot{r}^2 + r^2 B(r)}}, \ \ \ \dot{r} \equiv \der[r]{\phi}.
\end{equation}
Eqs.\ref{16} and \ref{22} imply
\begin{equation} \label{23}
\frac{\dot{r}^2}{r^4} + \frac{B(r)}{r^2} = \frac{1}{b^2}.
\end{equation}
Let us put
\begin{equation}
r = \frac{m}{u},
\end{equation}
$u$ being dimensionless. Then eq.\ref{23} takes the form
\begin{equation}\label{25}
\dot{u}^2 + u^2 B(u) = \frac{m^2}{b^2}.
\end{equation}
Taking the derivative of both sides with respect to $\phi$ one obtains the Lagrange equation in terms of $u$,
\begin{equation} \label{26}
\ddot{u} + u B(u) + \frac{u^2}{2} B'(u) = 0,\ B'(u) \equiv \der[B(u)]{u}. 
\end{equation}
Now, $B(u) = 1- 2u$ and \ref{26} reduces to \cite{12}, \cite{13}:
\begin{equation} \label{27}
\ddot{u} + u - 3u^2 =0
\end{equation}
Eq.\ref{27} is the Newton equation corresponding to the potential
\begin{equation}\label{28}
V(u) \equiv \frac{1}{2} u^2 B(u) = \frac{1}{2} u^2 - u^3.
\end{equation}
Eq.\ref{25} can be rewritten as
\begin{equation}\label{29}
\frac{1}{2} \dot{u}^2 + V(u) = \frac{m^2}{2b^2}.
\end{equation}
The "energy" $\frac{m^2}{2b^2}$ corresponds to the oscillatory regime provided $ \frac{m^2}{2b^2} < \frac{1}{54}$ and $u < \frac{1}{3}$ (Fig. \oldref{fig1}). For small $u$, $V(u) \sim u^2$ and, from eq.\ref{29}, $u^2 \lesssim \frac{m^2}{b^2}$. In the weak deflection limit, $b \gg m$, which gives $u \ll 1$. We are dealing with small nonlinear oscillations.

\begin{figure}[H] 
\centering
\begin{tikzpicture}[scale=14]
\begin{scope}[xscale = 1,yscale = 8]
\draw[->] (-0.2, 0) -- (0.6, 0) node[right] {$u$};
\draw[->] (0, -2/54) -- (0, 2/54) node[above] {$V$};
\draw[domain=-0.2:0.6,smooth,variable=\x] plot ({\x},{-\x*\x*\x+1/2*\x*\x});
\node at (0,0) [above right] {0};
\node at (1/2,0) [below] {$\frac{1}{2}$};
\draw[dotted] (1/3,0) node[below] {$\frac{1}{3}$}  -- (1/3, 1/54) ;
\draw[dotted] (0,1/54)node[left]  {$\frac{1}{54}$} -- (1/3, 1/54);
\end{scope}
\end{tikzpicture}
\caption{The potential corresponding to the Schwarzschield metric. Note that the value $u=\frac{1}{3}$ corresponds to the radius of circular light rays.} \label{fig1}
\end{figure}
As a second example of spherically symmetric metric we take the Reissner--Nordström one describing the charged black hole. It is given by eq.\ref{7} with
\begin{equation}
B(r) = A^{-1} (r) = 1 - \frac{2m}{r} + \frac{q^2}{r^2}, \ D(r)=1
\end{equation}
or
\begin{equation}
B(u) = A^{-1} (u) = 1 - 2u + \frac{q^2}{m^2} u^2.
\end{equation}
Eq.\ref{26} yields now
\begin{equation}\label{32}
\ddot{u} + u - 3u^2 + 2 \bracket{\frac{q}{m}}^2 u^3 = 0.
\end{equation}
The relevant potential reads (Fig \oldref{fig2})
\begin{equation}
\label{33}
V(u) = \frac{1}{2} u^2 - u^3 + \frac{1}{2} \bracket{\frac{q}{m}}^2 u^4.
\end{equation}
For the black hole to exist it is necessary that $\frac{q}{m} \leq 1$ (the equality corresponds to the extremal black hole).

\begin{figure}[H] 
\centering
\begin{tikzpicture}[scale=7]
\begin{scope}[xscale = 1,yscale = 7]
\draw[->] (-0.2, 0) -- (1.4, 0) node[right] {$u$};
\draw[->] (0, -0.1) -- (0, 0.1) node[above] {$V$};
\draw[domain=-0.2:1.18,smooth,variable=\x] plot ({\x},{0.5*\x*\x-\x*\x*\x + 0.45*\x*\x*\x*\x*\x});
\node at (0.57,0) [below] {$u_+$};
\node at (1.15,0) [below] {$u_-$};
\draw[dotted] (0,1/54) node[left]  {$\frac{m^2}{2b^2}$} -- (0.25, 1/54);
\end{scope}
\end{tikzpicture}
\caption{The potential corresponding to the Reissner--Nordström black-hole; $u_+$ and $u_-$ correspond to the outer and inner horizons, respectively.} \label{fig2}
\end{figure}

Again, the weak deflection limit $\frac{m}{b} \ll 1$ and $u<u_+$ imply small oscillations, $u\ll 1$.

\section{Axially symmetric metrics}
Let us now consider a more general case of axially symmetric metrics. We restrict ourselves to the motion in the equatorial plane $\theta = \frac{\pi}{2}$. Then the metric takes the form (in appropriate coordinates):
\begin{equation}
\dt[s]^2 = B(r) \dt[t]^2 - A(r) \dt[r]^2 - r^2 D(r) \dt[\phi]^2 + 2rF(r) \dt[t]\dt[\phi].
\end{equation}
Much of the previous discussion can be repeated here. The Lagrangian describing geodesic motion reads
\begin{align*}
L = \frac{1}{2} \left(B(r) \bracket{\der[t]{\sigma}}^2 - A(r) \bracket{\der[r]{\sigma}}^2 - r^2 D(r) \bracket{\der[\phi]{\sigma}}^2\right. \\
\left. + 2rF(r) \der[t]{\sigma} \der[\phi]{\sigma}\right)
\end{align*}
and yields the integrals of motion:
\begin{align}
E \equiv p_t = \partialder[\mathfrak{L}_0]{\dot{t}} &= B(r) \der[t]{\sigma} + r F(r) \der[\phi]{\sigma}\\
J \equiv p_\phi = \partialder[\mathfrak{L}_0]{\dot{\phi}}&= -r^2 D(r) \der[\phi]{\sigma} + r F(r) \der[t]{\sigma}\\
\end{align} 
The invariant impact parameter takes the form 
\begin{equation}
b = \abs{\frac{J}{E}} = \abs{\frac{-r^2 D(r) + r F(r) \der[t]{\phi}}{B(r) \der[t]{\phi} + r F(r)}}.
\end{equation} 
The Fermat principle \ref{4} is still valid so again by computing $\dt[x^0] \equiv \dt[t]$ from the equation $\dt[s]^2 = 0$ we find the variational principle for the light rays in the form:
\begin{align}
&\delta \int \mathfrak{L} \dt[\phi] = 0,\\
\label{40}
&\mathfrak{L} = \frac{-r F(r)}{B(r)} + \frac{1}{\sqrt{B(r)}} \der[l]{\phi},
\end{align}
with the length element of spatial distance
\begin{equation}
\dt[l]^2 = A(r) \dt[r]^2 + \bracket{r^2 D(r) + \frac{r^2 F^2(r)}{B(r)}} \dt[\phi]^2.
\end{equation}
$\mathfrak{L}$ does not depend explicitly on $\phi$ yielding $\bracket{\dot{r} \equiv \der[r]{\phi}}$
\begin{equation}\label{41}
\mathcal{E} = \dot{r} \partialder[\mathfrak{L}]{\dot{r}} - \mathfrak{L} = \frac{r F(r)}{B(r)} - \frac{r^2 D(r) + r^2 \frac{E^2(r)}{B(r)}}{\sqrt{B(r)} \sqrt{A(r) \dot{r}^2 + r^2 D(r) + \frac{F^2(r) r^2}{B(r)}}}
\end{equation}
as the conserved quantity. As in the spherically symmetric case the relation $\dt[t] = \mathfrak{L} \dt[\phi]$ implies \ref{16}.

By separating variables in eq.\ref{41} and integrating over $r$ one finds the counterpart of eq.\ref{19} for deflection angle. It appears to be quite complicated so we will not write it explicitly here.

The Lagrange equation which determines the trajectory $r = r(\phi)$, derived from $\mathfrak{L}$, involves $r$, $\dot{r}$ and $\ddot{r}$. Using \ref{41} one can eliminate $\dot{r}$ arriving at the equation of the form $\ddot{r} = g(r, \mathcal{E})$.

As an example of axially symmetric metric consider the Kerr black hole. The metric in the equatorial plane, when expressed in terms of $u$ coordinate, reads
\begin{align}\label{42}
B(u) &= 1 - 2u\\ \label{43}
A(u) &= \frac{1}{1-2u + \frac{k^2}{m^2} u^2}\\ \label{44}
D(u) &= 1 + \frac{k^2}{m^2} u^2 (1+2u)\\ \label{45}
F(u) &= - \frac{2k}{m} u^2
\end{align}
Using \ref{42} -- \ref{45} eq.\ref{41} can be converted into
\begin{equation}\label{46}
\dot{u}^2 = \frac{\bracket{1 - 2u + \frac{k^2}{m^2} u^2}^2}{\bracket{1 - 2u \bracket{1 - \frac{k}{\mathcal{E}}}}^2} \bracket{2 \bracket{1 - \frac{k}{\mathcal{E}}}^2 u^3 - \bracket{1 - \frac{k^2}{\mathcal{E}^2}} u^2 + \frac{m^2}{\mathcal{E}^2}}
\end{equation}
which coincides with eq.(11) from \cite{36}.

\section{Asymptotic trajectories and Lindstedt \\--Poincar\'e method}
We have seen that the light trajectories in the metric exhibiting high degree of symmetry are described by the simple non-linear equations: eq.\ref{27} for the Schwarzschild metric and eq.\ref{32} for the Reissner--Nordström one. They can be solved by writing down the first integrals, eq.\ref{25}, and separating variables. This results in the explicit expressions for light trajectories in terms of elliptic functions \cite{12}, \cite{13}. 

On the other hand, in the weak deflection regime $u \ll 1$, one can solve the relevant equations perturbatively in (the amplitude of) $u$.

However, one has to keep in mind that the straightforward application of perturbation theory produces the terms involving, apart from trigonometric functions, the polynomials in the angle $\phi$. This breaks periodicity and, consequently, the perturbative expansion is not valid uniformly. The reason is that in the recursive procedure there appear resonant terms producing unwanted contributions \cite{45, 46, 47, 48}. This can be cured by modifying the perturbative expansion according to the Lindstedt--Poincar\'e method \cite{45,46,47,48}: the consecutive terms (in particular, those entering the relation frequency -- amplitude) are determined from the condition that the resonance is absent. We sketch some details of Lindstedt--Poincar\'e method in the Appendix.

Let us start with the eq.\ref{27}. The Lindstedt--Poincar\'e method gives here, up to the second order in the amplitude,
\begin{align} \label{47}
u & = a \cos(\omega \phi) + \frac{a^2}{2} \bracket{3 - \cos(2 \omega \phi)} + ...\\ \label{48}
\omega & = 1 - \frac{15}{4} a^2 + ...
\end{align}
The above solution obeys the initial conditions:
\begin{align}\label{49}
u_0 &= u (\phi = 0) = a + a^2\\ \label{50}
\dot{u}_0 &= \dot{u}(\phi = 0) = 0
\end{align}
Any other solution may be obtained by the replacement $ \phi \to \phi - \phi_0$ with suitably adjusted $\phi_0$.
Eqs.\ref{28}, \ref{29}, \ref{49} and \ref{50} imply, up to the second order,
\begin{equation}\label{51}
\frac{m^2}{b^2} = u_0^2 - 2 u_0^3 \simeq a^2.
\end{equation}
Therefore, eqs.\ref{47}, \ref{48} may be rewritten as
\begin{align}\label{52}
u &= \frac{m}{b} \cos (\omega \phi) + \frac{m^2}{2 b^2} \bracket{3 - \cos (2 \omega \phi)}\\ \label{53}
\omega &= 1 - \frac{15}{4} \frac{m^2}{b^2}
\end{align}
Eq.\ref{52} provides the approximate form of light trajectory in the weak deflection limit. In order to find the value of deflection angle we solve first $u = 0$ for the asymptotic angle $\phi_\infty$ (cf. Fig. \oldref{fig3}):
\begin{equation}
\frac{m}{b} \cos (\omega \phi_\infty) + \frac{m^2}{b^2} \bracket{2 - \cos^2 (\omega \phi_\infty)} = 0
\end{equation} 
or, up to the second order,
\begin{equation}
\cos(\omega \phi_\infty) \simeq - \frac{2m}{b}.
\end{equation}
Let $\omega \phi_\infty \equiv \frac{\pi}{2} + \Delta$; then again up to the second order,
\begin{equation}\label{56}
\Delta \simeq \frac{2m}{b}
\end{equation}
On the other hand, the deflection angle $\delta$ is given by (see Fig. \oldref{fig3}):
\begin{equation}\label{57}
\delta = 2 \bracket{\phi_\infty - \frac{\pi}{2}} = \frac{\pi + 2\Delta}{\omega} - \pi.
\end{equation}
From eqs.\ref{53}, \ref{56} and \ref{57} one finds
\begin{equation}\label{58}
\delta \simeq \bracket{\pi + \frac{4m}{b}}\bracket{1 + \frac{15}{4} \bracket{\frac{m}{b}}^2} - \pi \simeq \frac{4m}{b} + \frac{15 \pi}{4} \bracket{\frac{m}{b}}^2
\end{equation}
which agrees with the value derived by other methods \cite{47a}. Let us stress that the second order correction to the Einstein original result follows, in this picture, from the frequency -- amplitude relation.

Now, one can include charge into our consideration. Again, eq.\ref{32} can be solved using Lindstedt--Poincar\'e algorithm. Evaluated to the second order the solution is still given by eq.\ref{52}. However, the relation frequency -- amplitude gets modified to
\begin{equation}
\omega = 1 + \bracket{\frac{3}{4} \bracket{\frac{q}{m}}^2 - \frac{15}{4}} a^2 + ...
\end{equation}
Following the same steps as in deriving eq.\ref{58} one finds
\begin{equation}\label{60}
\delta = \frac{4m}{b} + \frac{3 \pi}{4} \bracket{5 - \bracket{\frac{q}{m}}^2} \bracket{\frac{m}{b}}^2
\end{equation}
again in agreement with other findings \cite{7}. We see that the charge contribution results exclusively from amplitude -- frequency relation.

The above procedure can be extended to include higher order terms. First, consider the Schwarzschild metric. Then we have to solve eq.\ref{27}. Following the Lindstedt--Poincar\'e method we find easily $u$ to the third order in the amplitude,
\begin{equation}\label{61}
u = a \cos(\omega \phi) + \frac{a^2}{2} \bracket{3 - \cos(2 \omega \phi)} + \frac{3}{16} a^3 \cos(3 \omega \phi) + ...
\end{equation}
while eq.\ref{48} remains valid. Eq.\ref{61} implies 
\begin{equation}\label{62}
u_0 = a + a^2 + \frac{3}{16} a^3.
\end{equation}
Using this equation and and the first equality \ref{51} one obtains, to the third order in $\frac{m}{b}$,
\begin{equation}\label{63}
a = \frac{m}{b} + \frac{37}{16} \bracket{\frac{m}{b}}^3
\end{equation}
Eqs.\ref{61} and \ref{63} provide the approximate description of light ray trajectory up to the third order in $\bracket{\frac{m}{b}}$. On Figures \oldref{figSchwarzschild1}--\oldref{figSchwarzschild2} it is compared, for several values of $\frac{m}{b}$, with the exact solution.

\begin{figure}[H] 
\centering
\begin{tikzpicture}[scale=1]
\begin{scope}[xscale = 1,yscale = 1]
\begin{axis} [    
   	axis x line=center,
    axis y line=center,
    xlabel=$\phi$,
    ylabel=$u(\phi)$,
    xtick distance=0.25,
    ytick distance=0.03,
    xmin=0, xmax=2.1,  
    ymin=0, ymax=0.185,  
    width=12cm,   
    height=8cm,]
\addplot[only marks, mark=none, smooth] 
		table[x=Column1_135, y=Column2_135] {\mydata};
\addplot[only marks, mark=none, smooth, dashed] 
		table[x=Column1_135, y=Column3_135] {\mydata};
\node at (axis cs:0.5,0.06) {\Large{$\frac{m}{b}=0.135$}};
\end{axis}
\end{scope}
\end{tikzpicture}
\caption{Trajectory in the Schwarzschield metric for $\frac{m}{b} = 0.135$; solid line represents exact solution and the dashed line -- the approximate one.}\label{figSchwarzschild1}
\end{figure}

\begin{figure}[H] 
\centering
\begin{tikzpicture}[scale=1]
\begin{scope}[xscale = 1,yscale = 1]
\begin{axis} [    
   	axis x line=center,
    axis y line=center,
    xlabel=$\phi$,
    ylabel=$u(\phi)$,
    xtick distance=0.25,
    ytick distance=0.03,
    xmin=0, xmax=2.1,  
    ymin=0, ymax=0.14,  
    width=12cm,   
    height=8cm,]
\addplot[only marks, mark=none, smooth] 
		table[x=x01, y=Column2] {\mydata};
\addplot[only marks, mark=none, smooth, dashed] 
		table[x=x01, y=Column3] {\mydata};
\node at (axis cs:0.35,0.04) {\Large{$\frac{m}{b}=0.1$}};
\end{axis}
\end{scope}
\end{tikzpicture}
\caption{Trajectory in the Schwarzschield metric for $\frac{m}{b} = 0.1$; solid line represents exact solution and the dashed line -- the approximate one.}
\end{figure}

\begin{figure}[H] 
\centering
\begin{tikzpicture}[scale=1]
\begin{scope}[xscale = 1,yscale = 1]
\begin{axis} [    
   		axis x line=center,
    axis y line=center,
    xlabel=$\phi$,
    ylabel=$u(\phi)$,
    xtick distance=0.25,
    ytick distance=0.01,
    xmin=0, xmax=2,  
    ymin=0, ymax=0.061,  
    width=12cm,   
    height=8cm,]
\addplot[only marks, mark=none, smooth] 
		table[x=Column1_05, y=Column2_05] {\mydata};
\addplot[only marks, mark=none, smooth, dashed] 
		table[x=Column1_05, y=Column3_05] {\mydata};
\node at (axis cs:0.35,0.02) {\Large{$\frac{m}{b}=0.05$}};
\draw[dashed] (axis cs: 0,0.052) -- (axis cs: 0.25,0.052);
\draw[dashed] (axis cs: 0,0.054) -- (axis cs: 0.25,0.054);
\draw[dashed] (axis cs: 0.25,0.052) -- (axis cs: 0.25,0.054);
\draw[->] (axis cs: 0.25,0.053) -- (axis cs: 0.7,0.053); 
\node at (axis cs: 0.83,0.053) {Fig.\oldref{figSchwarzschild2}};
\end{axis}
\end{scope}
\end{tikzpicture}
\caption{Trajectory in the Schwarzschield metric for $\frac{m}{b} = 0.05$; solid line represents exact solution and the dashed line -- the approximate one.}
\end{figure}

\begin{figure}[H] 
\centering
\begin{tikzpicture}[scale=1]
\begin{scope}[xscale = 1,yscale = 1]
\begin{axis} [    
   		axis x line=center,
    axis y line=center,
    xlabel=$\phi$,
    ylabel=$u(\phi)$,
    xtick distance=0.1,
    ytick distance=0.0005,
    xmin=0, xmax=0.25,  
    ymin=0.052, ymax=0.054,  
    width=12cm,   
    height=8cm,]
\addplot[only marks, mark=none, smooth] 
		table[x=Column1_05, y=Column2_05] {\mydata};
\addplot[only marks, mark=none, smooth, dashed] 
		table[x=Column1_05, y=Column3_05] {\mydata};
\node at (axis cs:1.5,0.04) {\Large{$\frac{m}{b}=0.05$}};
\draw[dashed] (axis cs: 0,0.052) -- (axis cs: 0.25,0.052);
\draw[dashed] (axis cs: 0,0.054) -- (axis cs: 0.25,0.054);
\draw[dashed] (axis cs: 0.25,0.052) -- (axis cs: 0.25,0.054);
\end{axis}
\end{scope}
\end{tikzpicture}
\caption{Trajectory in the Schwarzschield metric for $\frac{m}{b} = 0.05$; solid line represents exact solution and the dashed line -- the approximate one.}\label{figSchwarzschild2}
\end{figure}

In particular, we would like to compute the deflection angle to the third order in $\frac{m}{b}$. According to the previous discussion we have to find $\phi_\infty$ by solving the equation $u=0$. To this end we put $y\equiv \cos(\omega\phi_\infty)$; then $u=0$ and eq.\ref{61} implies
\begin{equation}
    \label{65}
    \frac{3}{4} a^3 y^3 - a^2 y^2 + \bracket{1-\frac{9}{16} a^2} ay + 2a^2 = 0
\end{equation}
It is easy to see that $\phi_\infty$ differs from $\frac{\pi}{2}$ by $O(a)$ terms. Therefore, the relevant root $y$ of eq.\ref{65} is that of order $a \sim \frac{m}{b}$. We are looking for $y$ determined up to the order $\bracket{\frac{m}{b}}^3$. The structure of eq.\ref{65} clearly shows that, to this end, one has to include into eq.\ref{65} also the free term of fourth order. Therefore, in order to find $y$ to the third order from the equation $u=0$ one has to know $u(\phi)$ up to fourth order. Following the Lindstedt--Poincar\'e algorithm one finds in a straightforward way
\begin{align}
    \begin{split}
        u = &a \cos(\omega \phi) + \frac{a^2}{2} (3-\cos(2\omega\phi)) + \frac{3}{16} a^3 \cos(2\omega\phi)\\
        &+ \frac{57}{8} a^4 - \frac{59}{16} a^4 \cos(2\omega\phi) - \frac{a^4}{16} \cos(4\omega\phi) 
    \end{split}
\end{align}
On the other hand, it is sufficient to know $\omega$ up to the third order so we can still rely on eq.\ref{48}; also eq.\ref{63} remains valid. The corrected equation \ref{65} reads
\begin{align}
    \begin{split}
        -\frac{1}{2}a^4 y^4 &+ \frac{3}{4}a^3 y^3 - \bracket{\frac{55}{8} a^4 + a^2} y^2 + \bracket{1 - \frac{9}{16} a^2} ay\\
        &+\bracket{2a^2 + \frac{43}{4}a^4} =0
    \end{split}
\end{align}
with the solution, up to the third order, 
\begin{equation}
    y=-2a - \frac{63}{8} a^3
\end{equation}
Putting again
\begin{equation}
    \omega\phi_\infty \equiv \frac{\pi}{2} + \Delta
\end{equation}
one finds
\begin{equation}
    \sin \Delta = 2a+ \frac{63}{8} a^2
\end{equation}
or, using $\sin \Delta \simeq \Delta - \frac{1}{6} \Delta^3$,
\begin{equation}
    \Delta = 2a + \frac{221}{24} a^3
\end{equation}
Eqs.\ref{49} and \ref{58} imply now
\begin{equation}
    \delta \cong 4a + \frac{15\pi}{4} \bracket{\frac{m}{b}}^2 + \frac{128}{3}\bracket{\frac{m}{b}}^3
\end{equation}
in full agreement with \cite{7}.

The following remark is here in order. We have seen that to find the deflection angle in the third order approximation on has to know the $\phi$-independent part of $u$ to the fourth order. On the other hand we didn't encounter such problem when computing $\delta$ to the second order. This can be easily explained. It follows from the Lindstedt--Poincar\'e expansion that the even (odd) order terms involve even (odd) multiplicities of $\omega\phi$ as the arguments of cosine functions. Therefore, free terms appear only in the contributions of even orders. In particular, there is no contribution of third order to the constant term and one can compute the second order contribution to the deflection angle without addressing to the third order approximation. 

As a second example consider the Reissner--Nordström metric describing the charged black hole. The relevant equation \ref{32} can be again easily solved by Lindstedt--Poincar\'e method. To the fourth order we find
\begin{align}
    \begin{split}
    \label{74a}
        u = &a\cos(\omega\phi) + \frac{a^2}{2} (3-\cos(2\omega\phi)) +a^3 \bracket{\frac{3}{16} + \frac{1}{16} \bracket{\frac{q}{m}}^2} \cos(3\omega\phi) \\
        &+\bracket{\frac{57}{8} - \frac{15}{4} \bracket{\frac{q}{m}}^2} a^4 + \bracket{- \frac{59}{16}+\frac{31}{16} \bracket{\frac{q}{m}}^2} a^4 \cos(4\omega\phi)\\
        &- \bracket{\frac{1}{16} + \frac{1}{16}\bracket{\frac{q}{m}}^2} a^4 \cos(4\omega\phi)
    \end{split}
\end{align}
The frequency--amplitude relation is also modified,
\begin{equation}
    \omega = 1 + \bracket{- \frac{15}{4} + \frac{3}{4} \bracket{\frac{q}{m}}^2} a^2
\end{equation}
Again, as in the Schwarzschild metric case, $\dot{u} \equiv \dot{u} (\phi=0) = 0$ and eqs.\ref{29}, \ref{33} yield
\begin{equation}
    \label{76a}
    u_0^2 - 2u_0^3 + \bracket{\frac{q}{m}} u_0^4 = \bracket{\frac{m^2}{b^2}}
\end{equation}
Putting $\phi=0$ in eq.\ref{74a} one gets 
\begin{equation}
    \label{77}
    u_0 = a + a^2 + \bracket{\frac{3}{16} + \frac{1}{16} \bracket{\frac{q}{m}}^2} a^3 + \bracket{\frac{27}{8} - \frac{15}{8}\bracket{\frac{q}{m}}^2}a^4
\end{equation}
\ref{76a} and \ref{77} imply
\begin{equation}
    \label{78}
a=\frac{m}{b} + \bracket{\frac{37}{16} - \frac{9}{16} \bracket{\frac{q}{m}}^2} \bracket{\frac{m}{b}}^3    
\end{equation}
Eqs.\ref{74a} and \ref{78} determine the trajectory of light ray to the fourth order in $\frac{m}{b}$. The comparison between the exact solution and the approximate one  for several choices of $\frac{m}{b}$ parameter is presented on Figures \oldref{figReissner1}--\oldref{figReissner2}.

\begin{figure}[H] 
\centering
\begin{tikzpicture}[scale=1]
\begin{scope}[xscale = 1,yscale = 1]
\begin{axis} [    
    axis x line=center,
    axis y line=center,
    xlabel=$\phi$,
    ylabel=$u(\phi)$,
    xtick distance=0.2,
    ytick distance=0.02,
    xmin=0, xmax=2.12,  
    ymin=0, ymax=0.182,  
    width=12cm,   
    height=8cm,]
\addplot[only marks, mark=none, smooth] 
		table[x=phi, y=Reissner_mb0.135_qm0.5_exact] {\mydata};
\addplot[only marks, mark=none, smooth, dashed] 
		table[x=phi, y=Reissner_mb0.135_qm0.5_estim] {\mydata};
\node at (axis cs:0.6,0.08) {\Large{$\frac{m}{b}=0.135$, $\frac{q}{m}=0.5$}};
\end{axis}
\end{scope}
\end{tikzpicture}
\caption{Trajectory in the Reissner--Nordström metric for $\frac{m}{b} = 0.135$ and $\frac{q}{m}=0.5$; black line represents exact solution and red line -- the approximate one.}\label{figReissner1}
\end{figure}

\begin{figure}[H] 
\centering
\begin{tikzpicture}[scale=1]
\begin{scope}[xscale = 1,yscale = 1]
\begin{axis} [    
    axis x line=center,
    axis y line=center,
    xlabel=$\phi$,
    ylabel=$u(\phi)$,
    xtick distance=0.2,
    ytick distance=0.02,
    xmin=0, xmax=1.95,  
    ymin=0, ymax=0.13,  
    width=12cm,   
    height=8cm,]
\addplot[only marks, mark=none, smooth] 
		table[x=phi, y=Reissner_mb0.1_qm0.25_exact] {\mydata};
\addplot[only marks, mark=none, smooth, dashed] 
		table[x=phi, y=Reissner_mb0.1_qm0.25_estim] {\mydata};
\node at (axis cs:0.5,0.05) {\Large{$\frac{m}{b}=0.1$, $\frac{q}{m}=0.25$}};
\end{axis}
\end{scope}
\end{tikzpicture}
\caption{Trajectory in the Reissner--Nordström metric for $\frac{m}{b} = 0.1$ and $\frac{q}{m}=0.25$; black line represents exact solution and red line -- the approximate one.}
\end{figure}

\begin{figure}[H] 
\centering
\begin{tikzpicture}[scale=1]
\begin{scope}[xscale = 1,yscale = 1]
\begin{axis} [    
    axis x line=center,
    axis y line=center,
    xlabel=$\phi$,
    ylabel=$u(\phi)$,
    xtick distance=0.2,
    ytick distance=0.02,
    xmin=0, xmax=1.95,  
    ymin=0, ymax=0.13,  
    width=12cm,   
    height=8cm,]
\addplot[only marks, mark=none, smooth] 
		table[x=phi, y=Reissner_mb0.1_qm0.5_exact] {\mydata};
\addplot[only marks, mark=none, smooth, dashed] 
		table[x=phi, y=Reissner_mb0.1_qm0.5_estim] {\mydata};
\node at (axis cs:0.5,0.05) {\Large{$\frac{m}{b}=0.1$, $\frac{q}{m}=0.5$}};
\end{axis}
\end{scope}
\end{tikzpicture}
\caption{Trajectory in the Reissner--Nordström metric for $\frac{m}{b} = 0.1$ and $\frac{q}{m}=0.5$; black line represents exact solution and red line -- the approximate one.}
\end{figure}

\begin{figure}[H] 
\centering
\begin{tikzpicture}[scale=1]
\begin{scope}[xscale = 1,yscale = 1]
\begin{axis} [    
    axis x line=center,
    axis y line=center,
    xlabel=$\phi$,
    ylabel=$u(\phi)$,
    xtick distance=0.2,
    ytick distance=0.02,
    xmin=0, xmax=1.95,  
    ymin=0, ymax=0.13,  
    width=12cm,   
    height=8cm,]
\addplot[only marks, mark=none, smooth] 
		table[x=phi, y=Reissner_mb0.1_qm0.75_exact] {\mydata};
\addplot[only marks, mark=none, smooth, dashed] 
		table[x=phi, y=Reissner_mb0.1_qm0.75_estim] {\mydata};
\node at (axis cs:0.5,0.05) {\Large{$\frac{m}{b}=0.1$, $\frac{q}{m}=0.75$}};
\end{axis}
\end{scope}
\end{tikzpicture}
\caption{Trajectory in the Reissner--Nordström metric for $\frac{m}{b} = 0.1$ and $\frac{q}{m}=0.75$; black line represents exact solution and red line -- the approximate one.}
\end{figure}

\begin{figure}[H] 
\centering
\begin{tikzpicture}[scale=1]
\begin{scope}[xscale = 1,yscale = 1]
\begin{axis} [    
    axis x line=center,
    axis y line=center,
    xlabel=$\phi$,
    ylabel=$u(\phi)$,
    xtick distance=0.20,
    ytick distance=0.0050,
    xmin=0, xmax=1.8,  
    ymin=0, ymax=0.061,  
    width=12cm,   
    height=8cm,]
\addplot[only marks, mark=none, smooth] 
		table[x=phi, y=Reissner_mb0.05_qm0.5_exact] {\mydata};
\addplot[only marks, mark=none, smooth, dashed] 
		table[x=phi, y=Reissner_mb0.05_qm0.5_estim] {\mydata};
\node at (axis cs:0.4,0.02) {\Large{$\frac{m}{b}=0.05$, $\frac{q}{m}=0.5$}};
\draw[dashed] (axis cs: 0,0.052) -- (axis cs: 0.2,0.052);
\draw[dashed] (axis cs: 0,0.054) -- (axis cs: 0.2,0.054);
\draw[dashed] (axis cs: 0.2,0.052) -- (axis cs: 0.2,0.054);
\draw[->] (axis cs: 0.2,0.053) -- (axis cs: 0.7,0.053);
\node at (axis cs: 0.83,0.053) {Fig.\oldref{figReissner2}};
\end{axis}
\end{scope}
\end{tikzpicture}
\caption{Trajectory in the Reissner--Nordström metric for $\frac{m}{b} = 0.05$ and $\frac{q}{m}=0.5$; black line represents exact solution and red line -- the approximate one.}
\end{figure}

\begin{figure}[H] 
\centering
\begin{tikzpicture}[scale=1]
\begin{scope}[xscale = 1,yscale = 1]
\begin{axis} [    
    axis x line=center,
    axis y line=center,
    xlabel=$\phi$,
    ylabel=$u(\phi)$,
    xtick distance=0.025,
    ytick distance=0.00050,
    xmin=0, xmax=0.2,  
    ymin=0.052, ymax=0.054,  
    width=12cm,   
    height=8cm,]
\addplot[only marks, mark=none, smooth] 
		table[x=phi, y=Reissner_mb0.05_qm0.5_exact] {\mydata};
\addplot[only marks, mark=none, smooth, dashed] 
		table[x=phi, y=Reissner_mb0.05_qm0.5_estim] {\mydata};
\draw[dashed] (axis cs: 0,0.052) -- (axis cs: 0.2,0.052);
\draw[dashed] (axis cs: 0,0.054) -- (axis cs: 0.2,0.054);
\draw[dashed] (axis cs: 0.2,0.052) -- (axis cs: 0.2,0.054);
\node at (axis cs:0.4,0.03) {\Large{$\frac{m}{b}=0.05$, $\frac{q}{m}=0.5$}};
\end{axis}
\end{scope}
\end{tikzpicture}
\caption{Trajectory in the Reissner--Nordström metric for $\frac{m}{b} = 0.05$ and $\frac{q}{m}=0.5$; solid line represents exact solution and the dashed line -- the approximate one.}\label{figReissner2}
\end{figure}


In order to compute the deflection angle we solve the equation $u=0$ to the third order. It reads
\begin{align}
    \begin{split}
        \label{79}
        ay + a^2 (2-y^2) +a^3 \bracket{\frac{3}{16} + \frac{1}{16} \bracket{\frac{q}{m}}^2} \bracket{4y^3 - 3y}\\
        + a^4 \bracket{\frac{57}{8} - \frac{15}{4}\bracket{\frac{q}{m}}^2} + \bracket{-\frac{59}{16} + \frac{31}{16} \bracket{\frac{q}{m}}^2}\bracket{2y^2-1}\\
        -a^4 \bracket{\frac{1}{16}+\frac{1}{16} \bracket{\frac{q}{m}}^2} \bracket{8y^4 - 8y^2 +1}=0
    \end{split}
\end{align}
with the solution
\begin{equation}
    \label{80}
    y=-2a-\bracket{\frac{63}{8} - \frac{43}{8} \bracket{\frac{q}{m}}^2} a^3
\end{equation}
or, putting $y=-\sin\Delta \simeq -\Delta + \frac{1}{6} \Delta^3$,
\begin{equation}
    \label{81}
    \Delta =2a + \bracket{\frac{221}{24} - \frac{43}{8} \bracket{\frac{q}{m}}^2} a^3
\end{equation}
This leads to the following expression of deflection angle
\begin{equation}
    \label{82}
    \delta = 4a + \frac{15\pi}{4} a^2 - \frac{3\pi}{4} \bracket{\frac{q}{m}}^2 a^2 + \frac{401}{12} a^3 - \frac{55}{4} \bracket{\frac{q}{m}}^2 a^3
\end{equation}
or, by virtue of \ref{78},
\begin{equation}
    \label{83}
    \delta = 4 \frac{m}{b} + \bracket{\frac{15\pi}{4} - \frac{3\pi}{4} \bracket{\frac{q}{m}}^2} \bracket{\frac{m}{b}}^2 + \bracket{\frac{128}{3} - 16 \bracket{\frac{q}{m}}^2} \bracket{\frac{m}{b}}^3
\end{equation}
in full agreement with \cite{7}

Finally, consider the light rays propagation in equatorial plane of Kerr black hole. Eq.\ref{16} implies
\begin{equation}
    \mathcal{E} = s \abs{\mathcal{E}} = sb, \ \ \ s \equiv sgn \mathcal{E}
\end{equation}
Denoting 
\begin{equation}
    \alpha\equiv\frac{sk}{m},\hspace{4mm}  \beta \equiv \frac{m}{b} \ll 1
\end{equation}
one can rewrite \ref{47} in form
\begin{equation}
    \label{86}
    \dot{u}^2 = \frac{\bracket{1-2u +\alpha^2u^2}^2}{\bracket{1-2u(1-\alpha\beta}^2} \bracket{2(1-\alpha\beta)^2u^3 - \bracket{1-\alpha^2\beta^2}u^2 +\beta^2}
\end{equation}
By differentiating \ref{86} with respect to $\phi$ one obtains the equation of motion determining trajectory. It contains explicitly the small parameter $\beta$. The latter may be eliminated by using again the 'energy' integral \ref{86}. However, then the algebraic (non-rational) functions of $u$ and $\dot{u}$ appear. Obviously, the resulting equation coincides with the Lagrange equation obtained from the Lagrangian \ref{40}. The application of Lindstead--Poincar\'e method for the equation of motion in such a form may appear troublesome. In fact, one can try to expand the relevant algebraic functions in powers of $u$ and $\dot{u}$ but their quite complicated form would make the resulting equation difficult to manage.

We can, however, proceed differently. The condition $\dot{u}=0$ implies that one of the two brackets in the numerator on the right hand side of eq.\ref{86} vanishes. The vanishing of the first one implies $r$ being of order of the horizon radius. On the contrary, vanishing of the second bracket yields $u=O(\beta)$ and corresponds to the asymptotic region. Taking this into account we can expand the right hand side in powers of $\beta$ keeping in mind that $u$ is of order $\beta$. By terminating on some order and differentiating by $\phi$ one finds polynomial equation describing nonlinear oscillator. It involves explicitly the small expansion parameter $\beta$ but the Lindstedt--Poincar\'e method can still be applied; its essence is to eliminate systematically the resonant terms which is possible also in the case of $\beta$-dependent coefficients in equation of motion. One has to make only one necessary modification. The expansion parameter is $\beta$ instead of the amplitude $a$ of the motion. The relation between $a$ and $\beta$ is obtained by demanding that the value of 'energy' integral is $\frac{1}{2} \bracket{\frac{m}{b}}^2$ (cf. eq.\ref{29}). Now, we have a single parameter $\beta$ and the Lindstedt--Poincar\'e method produces uniquely defined successive terms. Therefore, we cannot take for granted that the energy of this particular solution equals $\frac{1}{2} \bracket{\frac{m}{b}}^2$ . To cure this at any step of perturbative expansion we add to the particular Lindstedt--Poincar\'e solution the general solution of free oscillator equation with the amplitude adjusted in such a way that the energy integrals equals, to a given order, $\frac{1}{2} \bracket{\frac{m}{b}}^2$.

In order to find the deflection angle to the third order in $\beta$ we have to know $u$ up to the fourth order. Therefore, eq.(86) should be expanded to fifth order. Then it takes the form
\begin{align}
    \begin{split}
        \dot{u}^2 = 2u^3 - u^2 +\beta^2 +3\alpha^2\beta^2 u^2 - 4 \alpha \beta^3 u - 2\alpha^2 u^4 + 6 \alpha^2 \beta^2u^3 - 8\alpha\beta^3u^2
    \end{split}
\end{align}
Differentiation over $\phi$ yields
\begin{equation}
    \label{87}
    \ddot{u}+u=3u^2+3\alpha^2\beta^2u - 2\alpha\beta^3 - 4 \alpha^2\beta^3 + 9\alpha^2\beta^2u^2 - 8\alpha\beta^3 u
\end{equation}
or, defining as in the Appendix, $\tau \equiv \omega \phi$ \ref{87} becomes
\begin{align}
    \begin{split}
    \label{89}
        \omega^2 \sder[u]{\tau} + u = 3u^2 + 3\alpha^2 \beta^2 u - 2 \alpha \beta^3 -4 \alpha^2 u^3 + 9 \alpha^2 \beta^2 u^2 - 8\alpha \beta^3 u
    \end{split}
\end{align}
Now, we expand both $u$ and $\omega$ in powers of $\beta$
\begin{align}
    \label{90}
    u &= u_1 + u_2 + u_3 + u_4\\
    \label{91}
    \omega &= 1 + \omega_1 +\omega_2 +\omega_3
\end{align}
One should keep in mind that the expansion \ref{91} concerns only the $\omega^2$ term appearing explicitly on the left hand side of eq.\ref{89}; $\omega$ entering the argument $\omega\phi$ of trigonometric functions is not expanded.

The algorithm providing the generalization of the Lindstedt--Poincar\'e method consists of the following steps: i) we substitute \ref{90} and \ref{91} into \ref{89} and write out the equation resulting from nullifying the coefficients in front of $\beta$, $\beta^2$, $\beta^3$, $\beta^4$ and $\beta^5$; ii) the resulting equations are solved recursively. Once the solution to order $n-1$ ($n=2,3,4$) is known the equation determining $u_n$ takes the form of that of forced harmonic oscillator with the force term determined by $u_1,...,u_{n-1}$; iii) the frequency $\omega$ is adjusted as to kill the resonant component of the force; iv) the particular solution for $u_n$ is found and supplied by the solution $A\cos \tau$ of free harmonic oscillator; v) $A$ is determined from the condition that \ref{87} is valid up to $n+1$st degree.

As a result we find:
\begin{align}
    u_1 = &\beta \cos\tau\\
    u_2 = &\frac{\beta^2}{2} (3-\cos{(2\tau)})\\
    u_3 = &\bracket{\frac{37}{16} + \frac{3\alpha^2}{8}} \beta^3 \cos\tau - 2 \alpha \beta^3 + \bracket{\frac{3}{16} + \frac{\alpha^2}{8}} \beta^3 \cos(3\tau)\\
    \begin{split}
    u_4 = &-10 \alpha \beta^4 \cos \tau + \bracket{\frac{225}{16} + \frac{21 \alpha^2}{8}} \beta^4 + \bracket{-6 + \frac{\alpha^2}{2}} \cos (2\tau) \\
    &- \bracket{\frac{1}{16} + \frac{\alpha^2}{8}} \cos(4\tau)
    \end{split}
\end{align}
\begin{align}
    \omega_1 = 0,\hspace{3mm} \omega_2= -\frac{15}{4}\beta^2,\hspace{3mm} \omega_3=10\alpha\beta^3
\end{align}
Coming back to the original notation we find the following equation for light ray trajectory up to the $\bracket{\frac{m}{b}}^4$ order:
\begin{align}
    \begin{split}
    \label{97}
        u = &\frac{m}{b} \cos(\omega\phi) + \frac{1}{2}\bracket{\frac{m}{b}}^2 (3-\cos(2\omega\phi)) - 2\ \frac{sk}{m} \bracket{\frac{m}{b}}^3 \\
        &+ \bracket{\frac{3}{16} + \frac{1}{8} \bracket{\frac{k}{m}}^2}\bracket{\frac{m}{b}}^3 \cos(3\omega\phi) - 10\ \frac{sk}{m} \bracket{\frac{m}{b}}^4 \cos(\omega\phi)\\
        &+\bracket{\frac{225}{16} + \frac{21}{8} \bracket{\frac{k}{m}}^2}\bracket{\frac{m}{b}}^4 + \bracket{-6 + \frac{1}{2}\bracket{\frac{k}{m}}^2} \cos(2\omega\phi) \\
        &- \bracket{\frac{1}{16} + \frac{1}{8} \bracket{\frac{k}{m}}^2}\bracket{\frac{m}{b}}^4 \cos(4\omega\phi)
    \end{split}
\end{align}
\begin{equation}
    \omega = 1 - \frac{15}{4} \bracket{\frac{m}{b}}^2 + 10\ \frac{sk}{m}\bracket{\frac{m}{b}}^3
\end{equation}

Now we can repeat steps leading to the third order expression for deflection angle.\\
First, u=0 gives
\begin{align}
    \begin{split}
        \label{99}
        &\beta y + \beta^2 (2-y^2) + \bracket{\frac{37}{16} + \frac{3\alpha^2}{8}} \beta^3 y - 2 \alpha \beta^3 + \bracket{\frac{3}{16} + \frac{\alpha^2}{8}}\bracket{4y^3-3y}\\
        &-10\ \alpha \beta^4 y + \bracket{\frac{225}{16} + \frac{21 \alpha^2}{8}} \beta^4 + \bracket{-6 + \frac{\alpha^2}{2}} \beta^4\ (2y^2-1) \\
        &- \bracket{\frac{1}{16} + \frac{\alpha^2}{8}} \beta^4 (8y^4-8y^2+1)=0
    \end{split}
\end{align}
yielding
\begin{equation}
    y=-2\beta + 2 \alpha\beta^2-\bracket{\frac{25}{2} + 2\alpha^2} \beta^3
\end{equation}
or
\begin{equation}
    \Delta = 2\beta - 2 \alpha\beta^2 + \bracket{\frac{83}{6} + 2\alpha^2} \beta^3
\end{equation}
Due to $\delta=\frac{\pi+2\Delta}{\omega} - \pi$ we get
\begin{equation}
    \delta=4\ \frac{m}{b} + \bracket{\frac{15\pi}{4}-4\ \frac{sk}{m}}\bracket{\frac{m}{b}}^2 + \bracket{\frac{128}{3} -10\pi \ \frac{sk}{m} + 4\bracket{\frac{k}{m}}^2} \bracket{\frac{m}{b}}^3
\end{equation}
in agreement with the result obtained by other method \cite{36}.

\section{The advantage of Lindstedt--Poincar\'e approach}

We have seen that the description of light rays in weak deflection regime reduces, at least in the cases of highly symmetric metric, to that of small nonlinear oscillations (cf. eqs. \ref{27} and \ref{32}) . Therefore, one can apply the well-developed perturbative methods. The naive perturbative expansion produces the sum of terms, each being the product of some polynomial in independent variable (here the azimuthal angle $\phi$) and the trigonometric function ($\sin$ and/or $\cos$) of multiple $\omega_0 \phi$, $\omega_0$ being the unperturbed frequency. The polynomial coefficients emerge due to resonance phenomenon: the "force" composed from lower order contributions typically contains the resonant terms. Obviously, a naive expansion is sufficient to compute, to a given order, the overall characteristics of the trajectories like deflection angle \cite{9}, \cite{54}, \cite{55}. However, it fails to reflect the properties of trajectory as a function of azimuthal angle. To see this let us consider the simplest of Schwarzschield metric. It is clearly seen from eqs. \ref{27}--\ref{29} and Fig. 2. that the $u$-motion is periodic while the naive approximation in not. On the contrary, the Lindstedt--Poincar\'e expansion is periodic to any order, the period being determined perturbatively. Therefore, it reflects correctly the properties of exact trajectory.

Consider the trajectory for some (small) value of $\frac{m}{b}$. The radial motion is described by the trajectory in potential well depicted on Fig 2. The physically meaningful region corresponds to $u>0$. Suppose we start the right neighbourhood of $u=0$ and move to the right. At $\phi=0$ $u$ bounces off the potential barrier and moves towards 0; this corresponds to a single physical trajectory. Once $u$ leaves the right semiaxis the trajectory disappears. When $u$ bounces off the left potential barrier and reaches the origin new physical trajectory emerges and the cycle repeats. We conclude that the light trajectories corresponding to a fixed invariant impact parameter naturally decompose into disjoint sets, the trajectories belonging to a given set emerge from a particular nonlinear oscillations.

The above property can be easily illustrated using second order approximation given by eq. \ref{53}. Eqs. \ref{56} and \ref{52} imply that the azimuthal angles corresponding to the set containing the trajectory \ref{53} takes the values belonging to the intervals
\begin{align}
\nonumber
\phi  \in & \left(\bracket{ 2n - \frac{1}{2} } \pi \bracket{ 1 + \frac{15}{4} \bracket{\frac{m}{b}}^2 } - \frac{2m}{b} ,  \right.
\\
&\left. \bracket{2n + \frac{1}{2}} \pi \bracket{ 1 + \frac{15}{4} \bracket{\frac{m}{b}}^2 } + \frac{2m}{b} \right) 
\\
\nonumber
& n = 0,\pm 1, \pm 2,...&
\end{align}
The sets of related trajectories are finite or infinite depending on whatever $\bracket{\frac{m}{b}}^2$ is rational or rational, respectively. This is, however, an artefact of the approximation considered. An example of the family of related trajectories is given on Fig. \ref{fig14}.


\begin{figure}[H] 
\centering
\begin{tikzpicture}[scale=1]
\begin{scope}[xscale = 1,yscale = 1]
\begin{axis} [    
    axis x line=center,
    axis y line=center,
    xtick distance=5,
    ytick distance=5,
    xmin=-19, xmax=24,  
    ymin=-19, ymax=24,  
    width=12cm,   
    height=12cm,]
\addplot[only marks, mark=none, smooth] 
		table[x=iks1, y=igrek1] {\mydata};
\addplot[only marks, mark=none, smooth] 
		table[x=iks2, y=igrek2] {\mydata};	
\addplot[only marks, mark=none, smooth] 
		table[x=iks3, y=igrek3] {\mydata};
\addplot[only marks, mark=none, smooth] 
		table[x=iks4, y=igrek4] {\mydata};	
\addplot[only marks, mark=none, smooth] 
		table[x=iks5, y=igrek5] {\mydata};
\addplot[only marks, mark=none, smooth] 
		table[x=iks6, y=igrek6] {\mydata};
\addplot[only marks, mark=none, smooth] 
		table[x=iks7, y=igrek7] {\mydata};	
\addplot[only marks, mark=none, smooth] 
		table[x=iks8, y=igrek8] {\mydata};
\addplot[only marks, mark=none, smooth] 
		table[x=iks9, y=igrek9] {\mydata};	
\addplot[only marks, mark=none, smooth] 
		table[x=iks10, y=igrek10] {\mydata};	
\addplot[only marks, mark=none, smooth] 
		table[x=iks11, y=igrek11] {\mydata};
\addplot[only marks, mark=none, smooth] 
		table[x=iks12, y=igrek12] {\mydata};	
\addplot[only marks, mark=none, smooth] 
		table[x=iks13, y=igrek13] {\mydata};
\addplot[only marks, mark=none, smooth] 
		table[x=iks14, y=igrek14] {\mydata};	
\addplot[only marks, mark=none, smooth] 
		table[x=iks15, y=igrek15] {\mydata};								
\end{axis}
\end{scope}
\end{tikzpicture}
\caption{Family of related trajectories for $\frac{m}{b}=0.1$. \label{fig14} } 
\end{figure}

Due to the rotational symmetry the trajectories under consideration (with the fixed impact parameter) are generated from a single trajectory by rotations around orthogonal axis. On the level of $u$-oscillations this corresponds to the generation of all oscillations of a given energy by "time" translation from a single oscillation. This important feature is properly captured by Lindstedt--Poincar\'e expansion. Moreover, contrary to the naive perturbative expansion, Lindstedt--Poincar\'e approach preserves, at each step, the basic properties of exact solution, in particular its periodicity. The latter results in decomposition of the set of light trajectories into disjoint subsets of trajectories corresponding to a single oscillation mode. 

\section{Conclusions}

Fermat's principle provides a very convenient method for studying the propagation of light in curved spacetime in the geometric optic approximation. The first advantage is that it can be formulated in extremely simple way. One solves the light cone condition $\dt[s]^2=0$ with respect to $\dt[x]^0$ yielding the integrand of the relevant variational principle. In this way one obtains the appropriate Lagrangian; the corresponding Lagrange equations represent the infinitesimal form of the Fermat principle. Once they are solved the space trajectory of the light rays is determined. The time variable $x^0$ can be obtained by integrating the Lagrange function along the trajectory; in other words, $x^0$ may be identified with the action variable. 

Secondly, the resulting Lagrangian formalism is degenerate --- it is reparametrization invariant. There is a freedom in the choice of evolution parameter. In particular, we can choose (at least locally) one of the space coordinates $x^i$ thereby reducing the number of variables under consideration. 

Originally, Fermat's principle has been formulated for time-independent gravitational fields, both static and stationary. However, it appeared \cite{48,49,50,51} that the Hamiltonian and Lagrangian form of Fermat's principle is possible also for arbitrary gravitational field. The price to be payed is that one has to use the generalized version of Lagrangian/Hamiltonian formalism developed by Herglotz \cite{52}.

In the limit of large impact parameter one can try to solve the equations resulting from Fermat's principle perturbatively. Since we are dealing with the equations formally describing the (set of) nonlinear oscillators, the perturbation expansion should be modified to avoid the appearance of resonant terms. The relevant Lindstedt--Poincar\'e formalism provides uniform approximation to exact trajectories. As we have shown, it can be used to compute the deflection angle in the weak deflection regime. The actual calculations involve only algebraic operations; neither any integration nor Fourier expansion of elliptic functions are necessary. 

The partial summation involved in the Lindstedt--Poincar\'e algorithm allows also to reflect correctly the global properties of the ray trajectories, in particular --- their periodicity which, in physical terms, implies that the set of all trajectories with a given impact parameter decomposes naturally into disjoint subsets, each being generated by a particular oscillatory solution.

We have considered above only the integrable cases, where the analytic solutions are available. However, we believe the Lindstedt--Poincar\'e method is more flexible and can be applied even if the geodesic equations are not integrable. However, even in the integrable case it reveals interesting properties. For example, in the simplest case of Schwarzschild metric the first correction to the Einstein expression  for deflection angle comes from the frequency -- amplitude dependence characteristic for nonlinear oscillators. The same holds true for the charged black hole.

In the subsequent publications we will dealing with more complicated cases where analytical solutions are not available. 

\bibliographystyle{ieeetr}
\bibliography{lindstedt_poincare}
\appendix

\section{The Lindstedt--Poincar\'e method \cite{43, 47, 48, 46}}
Consider small non-linear oscillations described by the equation $\bracket{\ddot{u} \equiv \sder[u]{\phi}}$,
\begin{equation}\label{A1}
\ddot{u} + \sum\limits_{n=1}^N \alpha_n u^n =0, \qquad \alpha_1 \equiv \omega_0^2
\end{equation}
Assuming $u \ll 1$ one can solve \ref{A1} perturbatively. However, the straightforward application of perturbative expansion yields the so-called secular terms of the form: polynomial in $\phi$ times trigonometric function. Because of them the perturbative expansion is not uniformly valid as $\abs{\phi}$ increases. 

In order to get rid of secular terms one introduces new independent variable
\begin{equation} \label{A2}
\tau \equiv \omega \phi
\end{equation}
and expands both $\omega$ and $u$ in the powers of oscillation amplitude,
\begin{equation} \label{A3}
\omega = \omega_0 + \omega_1 + \omega_2 + ...
\end{equation}
\begin{equation} \label{A4}
u = u_1 + u_2 + u_3 + ...
\end{equation}
Inserting \ref{A2} -- \ref{A4} into \ref{A1} one arrives at the chain of equations
\begin{equation} \label{A5}
\sder[u_1]{\tau} + u_1 =0
\end{equation}
\begin{equation}\label{A6}
\omega_0^2 \bracket{\sder[u_2]{\tau} + u_2} = -2 \omega_0 \omega_1 \sder[u_1]{\tau} - \alpha_2 u_1^2
\end{equation}
\begin{align} \label{A7}
\omega_0^2 \bracket{\sder[u_3]{\tau} + u_3} = -2 \omega_0 \omega_1 \sder[u_2]{\tau} - 2 \alpha_1 u_1 u_2 +\\ \notag
- \bracket{\omega_1^2 + 2 \omega_0 \omega_2} \sder[u_1]{\tau} - \alpha_3 u_1^3
\end{align}
etc.
\ref{A5} gives
\begin{equation} \label{A8}
u_1 = a \cos \tau;
\end{equation}
the expansion we are considering is, in fact, the one in the power of $a$. Substituting \ref{A8} into \ref{A6} one finds
\begin{equation}\label{A9}
\omega_0^2 \bracket{\sder[u_2]{\tau} + u_2} = 2 \omega_0\omega_1 a \cos \tau - \frac{1}{2}\ \alpha_2 a^2 \bracket{1 + \cos 2 \tau}
\end{equation}
The first term on the right hand side is resonant, leading to the contribution of the form $\tau \sin \tau$ which we don't want. Therefore, we put $\omega_1 = 0$ and solve \ref{A9} for $u_2$:
\begin{equation}\label{A10}
u_2 = \frac{- \alpha_2 a^2}{2 \omega_0^2} \bracket{1 - \frac{1}{3} \cos 2 \tau}
\end{equation}
As a next step we insert \ref{A8} and \ref{A10} into \ref{A7}. The resonant term is then absent provided
\begin{equation}
\omega_2 = \bracket{\frac{9 \alpha_3 \omega_0^2 - 10 \alpha_2^2}{24 \omega_0^3}} a^2
\end{equation} 
and one can solve \ref{A7} for $u_3$ which appears to be proportional to $\cos 3 \tau$. This procedure can be continued with higher order terms. The consecutive terms in the expansion \ref{A3} are obtained by demanding that the resonant contributions to \ref{A4} are absent.
\end{document}